\documentclass{aa}
\usepackage{epsfig,times}
\usepackage{amssymb}
\usepackage{graphicx}
\begin{document}

\title{Joule heating in high magnetic field pulsars}
\author{ V.~Urpin \inst{1,2} and D.~Konenkov \inst{2}}
\offprints{Vadim.Urpin@uv.es}
\institute{$^{1)}$ Departament de F\'{\i}sica Aplicada, Universitat d'Alacant,
           Ap. Correus 99, 03080 Alacant, Spain \\
           $^{2)}$ A.F.Ioffe Institute of Physics and Technology and Isaac 
           Newton Institute of Chile, Branch in St.Petersburg, 194021 
           St.Petersburg, Russia}
\date{today}
\abstract
{} 
{We study the efficiency of Joule heating in the crustal layers of young
neutron stars.}
{The induction equation is solved numerically for a realistic crust model
with a conductivity that depends on the density and temperature.}
{We show that dissipation of the magnetic field is highly inhomogeneous in
the crust with much faster dissipation in relatively low density layers. In 
young neutron stars, the rate of Joule heating in the crust can  
exceed the standard luminosity of non-magnetic stars and can even be 
comparable to the luminosity of magnetars. }
{The results of calculations are compared with the available observational
data. We argued that the crustal field model can well account for the data 
on the surface temperature and magnetic field of young neutron stars.}
\keywords{MHD - stars: neutron - stars: magnetic fields - pulsars: general}

\titlerunning{Joule heating in high magnetic field pulsars}
\authorrunning{V.Urpin and D.Konenkov}

\maketitle

\section{Introduction}

The influence of the magnetic field on the thermal history of neutron stars
is twofold. On one hand, the field changes the character of heat transport 
providing some anisotropy to kinetic processes. Such anisotropic transport 
can result in a characteristic temperature difference between the poles and
equator caused by large-scale inhomogeneities of the magnetic field (see, 
e.g., Potekhin et al. 2003; Perez-Azorin et al. 2006). On the other hand, 
dissipation of the magnetic field provides an additional heating mechanism 
that can substantially alter the cooling evolution (Miralles et al. 1998). 
Even more or less standard for radiopulsars magnetic fields $\sim 10^{12} -
10^{13}$ G can maintain the surface temperature as high as $\sim 10^{5}$ K 
for $\sim 10-100$ Myr which is substantially higher than predicted by
various cooling models. A general relativistic treatment of Joule heating 
in neutron stars undertaken by Page et al. (2000) led to results very 
similar to those of Miralles et al. (1998) since, as it was argued by Konar 
(2002), general relativistic corrections do not significantly influence 
the magnetic field decay in neutron stars. 

The internal magnetic fields in neutron stars, however, can be substantially 
stronger than the ``standard'' surface field of radiopulsars, and they can 
produce much more efficient heating. The nature of pulsar magnetism is still 
the subject of debate. One of the possibilities is associated with the 
turbulent dynamo action that can amplify the magnetic field during the first 
$\sim 30-40$ s of a neutron star's life when the star is subject  
hydrodynamic instabilities (Thompson \& Duncan 1993; Bonanno et al. 2003). 
Turbulent motion caused by instabilities in combination with rapid rotation 
that seems to be almost inevitable in protoneutron stars make a turbulent 
dynamo one of the most plausible mechanisms of pulsar magnetism. The 
characteristic feature of a turbulent dynamo in stars is that generation of 
the large-scale field is always accompanied by generation of small-scale 
fields that are approximately in equipartition with the turbulent motion. 
Therefore, if the field of pulsars is of dynamo origin, it should have a very 
complex geometry when dynamo action stops (see Urpin \& Gil 2004, Bonanno et 
al. 2005, 2006). Magnetic fields of various scales will be frozen into the 
crust that begins to form soon after the end of the unstable stage and then 
will decay due to the finite crustal resistivity.   

Miralles et al.(1998) have considered the effect of Joule heating in old 
neutron stars with an age $t > 10^6$ yrs, when the surface temperature is 
low and dissipation of even standard pulsar magnetic fields can change the 
thermal evolution substantially. In this paper, we treat the effect of 
dissipation during the early evolution of neutron stars ($t < 10^5$ yrs), 
when neutron stars are much hotter. The crustal conductivity depends on the 
temperature and can be relatively low in hot neutron stars. Due to this, the 
field decay is rapid in young neutron stars with an age $\sim 10^4-10^5$ yrs.
This type of behaviour has been well known since the paper by Urpin \& 
Muslimov (1992). The fast decay has been studied in detail by a number of 
authors (see, e.g., Urpin \& Konenkov 1997; Sengupta 1998; Mitra et al. 1999, 
etc.), and was rediscovered with no citation in recent calculations by Pons 
\& Geppert (2007). However, even such efficient field decay during the initial 
stage can influence the thermal evolution only if the magnetic field is 
sufficiently strong and, therefore, our study primarily addresses high 
magnetic field pulsars.

\section{The internal magnetic fields in pulsars}

The unstable stage lasts only $\sim 30-40$ s in proto-neutron stars, but 
this is sufficient for dynamo action to generate a strong magnetic field 
because the turbulent motion is very fast. A turbulent dynamo can generate a 
large-scale magnetic field only if the star rotates rapidly and the rotation 
period $P_0$ is shorter than some ``critical'' value $P_{cr}$ ($\sim 1$ s in 
the simplest model). Most proto-neutron stars likely rotate faster and, as 
a result, a dynamo should operate in these objects (Bonanno et al. 2003, 
2005). The dynamo generates both the toroidal field that is essentially 
internal and vanishes at the surface and the poloidal field that is 
non-vanishing outside the star. The poloidal field is responsible for pulsar 
activity. Typically, the generated toroidal field is comparable to the 
poloidal one if the proto-neutron star rotates rigidly, but it can be 
substantially stronger if rotation is differential. For example, the toroidal 
field is $\sim 20-60$ times stronger than the poloidal one if rotation at the 
center is three times faster than that at the surface equator. The saturation 
field generated by the dynamo can be estimated by making use of the simplest 
model of $\alpha$-quenching (Bonanno et al. 2006). This yields for the 
generated large-scale field
\begin{equation}
B_{0} \approx B_{eq} \sqrt{P_{cr}/P_{0}-1} ,
\end{equation}
where $B_{eq}$ is the equipartition small-scale field, $B_{eq} \approx 4 \pi 
\rho v_{T}^{2}$ where $v_T$ is the turbulent velocity. The equipartition 
field varies during the unstable stage, rising rapidly soon after collapse 
and then decreasing when the temperature and lepton gradients are smoothed. 
At the peak, $B_{eq}\sim 10^{16}$ G in the convective zone (Thompson \& Duncan
1993) and $B_{eq} \sim (1-3) \times 10^{14}$ G in the neutron-finger unstable 
zone (Urpin \& Gil 2004). However, $B_{eq}$ decreases and becomes smaller 
than the peak value as the protoneutron star cools down. The dynamo operates 
while the quasi-steady condition $\tau_{cool} \gg \tau$ is fulfilled with 
$\tau_{cool}$ being the cooling timescale and $\tau$ the turnover time of 
turbulence. The final strength of a generated small-scale magnetic field is 
approximately equal to $B_{eq}$ at the moment when the quasi-steady condition 
breaks down and $\tau \sim \tau_{cool}$. If $\tau \sim \tau_{cool}$, we have 
$v_{T} \sim \pi \ell_{T} /\tau_{cool}$ where $\ell_T$ is the turbulent 
lengthscale. Then, 
\begin{equation}
B_{eq} \sim  \sqrt{4 \pi \rho} v_{T} \sim 
\pi \sqrt{4 \pi \rho} \ell_{T} \tau_{cool}^{-1}.
\end{equation}
Estimate (2) yields $B_{eq} \sim (1-3) \times 10^{13}$ G for the maximum
possible lengthscale $\ell_{T} \sim 1-3$ km and $\tau_{cool} \sim$ few 
seconds. 

Substituting estimate (2) into expression (1), we obtain
\begin{equation}
B_{0} \approx \frac{ \sqrt{4 \pi^3 \rho} \ell_{T}}{\tau_{cool}} \;
\sqrt{ \frac{P_{cr}}{P_{0}} -1} \sim (1-3) \times 10^{13} 
\sqrt{ \frac{P_{cr}}{P_{0}} - 1} \;\; {\rm G}.
\end{equation}
The shortest possible period is likely $\sim 1$ ms and, hence, the strongest 
internal magnetic field generated by a dynamo can reach $\sim (0.3-1) \times 
10^{15}$ G. The field at the surface is usually weaker than the internal field
generated by a dynamo (Bonanno et al. 2006). Note that generation is more 
efficient in the outer layers because the parameter $\alpha$ that determines 
the efficiency of the dynamo is proportional to $\Omega$ and the density 
gradient (see, e.g., R\"udiger \& Kitchatinov 1993) but the latter is maximal 
in the outer layers.

If $P_{0}$ is longer than $P_{cr}$, then the large-scale dynamo does not 
operate in proto-neutron stars but the small-scale dynamo is still efficient. 
One can expect that such neutron stars have only small-scale fields with the 
strength $B_{eq} \sim (1-3) \times 10^{13}$ G and have no dipole field. 
Likely, slow rotation with $P_0 > P_{cr}$ is rather difficult to achieve in
proto-neutron stars, and the number of such exotic objects is small. 

It is difficult to predict the final disposition of the magnetic field after 
turbulent motion stops. The magnetic configuration generated by the dynamo 
can be unstable, for example, if it contains a strong toroidal component. MHD 
instabilities can lead to a rapid evolution on the Alfv\'en time scale and 
form a configuration with comparable poloidal and toroidal fields (see, 
e.g., Braithwaite \& Spruit 2006; Braithwaite \& Nordlund 2006). Therefore, 
it is unlikely that the toroidal field is substantially stronger than the 
poloidal one in neutron stars. When convection is exhausted, both large- and 
small-scale fields evolve under the influence of ohmic dissipation and 
buoyancy. Likely, formation of the crust provides the most important 
influence on the evolution of magnetic fields at this stage. Soon after 
convection stops, the neutron star cools down to the internal temperature 
$\sim (1-3) \times 10^{10}$ K at which neutrons and protons can form nuclei 
and clusters in matter with a density $\sim 10^{14}$ g/cm$^3$. When 
the temperature decreases progressively, nuclei can be formed at a lower 
density as well. Nuclear composition at high density is subject to many 
uncertainties and depends generally on the pre-history of the neutron star. 
For instance, the composition of the ``ground state'' matter (Negele \& 
Vautherin 1973) differs noticeably from that of the ``accreted'' matter 
(Haensel \& Zdunik 1990). The Coulomb interactions of nuclei leads to 
crystallization that occurs when the ion coupling parameter $\Gamma = 
Z^2 e^2/ a k_B T$ reaches the critical value $\Gamma_m \approx 170$ 
(Slattery et al. 1980); $a=( 3/4 \pi n_i)^{1/3}$ is the mean inter-ion 
distance, $n_i$ and $Z$ are the number density and charge number of ions, 
respectively; $T$ is the temperature, and $k_B$ is the Boltzmann constant. 
Then, the crystallization temperature is 
\begin{equation}
T_m \approx 1.3 \times 10^7 Z^{5/3} \mu_e^{-1/3} \rho_{12}^{1/3}
\;\; {\rm K}, 
\end{equation}
where $\mu_e$ is the number of baryons per electron, and $\rho_{12} =
\rho/10^{12}$ g/cm$^3$. For $\rho \sim 10^{13}-10^{14}$ g/cm$^3$, the 
crystallization temperature is $\sim 10^{10}$ K. Therefore, the crust 
formation starts almost immediately after the end of the convective phase 
(or even before the end), and the magnetic field generated by the
dynamo should be frozen into the crust.

\section{Joule heating in the crust}

Likely, the evolution of the magnetic field in the crust of hot neutron 
stars is mainly determined by the ohmic dissipation, and Joule heating has 
an important influence on the thermal evolution. The rate of Joule heating 
depends both on the geometry of the magnetic field and the conductive 
properties of the crust. In a strong magnetic field, the ohmic dissipation 
can be accompanied by Hall currents (Shalybkov \& Urpin 1997, Hollerbach 
\& R\"udiger 2002, 2004). These currents are non-dissipative since they do 
not contribute directly to the rate of Joule heating. However, the 
Hall currents couple different modes and alter the magnetic configuration by 
redistributing the energy among modes. In this way, the Hall currents can 
indirectly affecr the rate of dissipation. Note also that, because of the
non-linear nature of the Hall currents, they can generate magnetic features 
at the stellar surface with a smaller lengthscale than the background 
magnetic field. This has been known since the paper by Naito \& Kojima (1994) 
(see also Muslimov 1994, Shalybkov \& Urpin 1997) who argued that some 
neutron stars may have strong disordered magnetic fields at their surface due 
to the Hall effect. The same conclusion recently was reached by Pons 
\& Geppert (2007).   

The influence of the Hall effect on the magnetic field is characterized by 
the Hall parameter, $\omega_B \tau$, where $\omega_B = e B / m_{*} c$ is the 
gyrofrequency of electrons and $\tau$ is their relaxation time, $m_{*}$ is 
the electron effective mass. The Hall effect is unimportant if $\omega_B
\tau \leq 1$, but it can influence the magnetic evolution if the field is 
sufficiently strong and $\omega_B \tau \gg 1$. Indeed, numerical simulations 
(Shalybkov \& Urpin 1997, H\"ollerbach \& R\"udiger 2002, 2004) indicate 
that some acceleration of the decay can occur, particularly if the Hall 
parameter is very large, $\omega_B \tau > 200$. This conclusion, however,
turns out to be sensitive to the initial magnetic geometry. For a purely 
toroidal field, the Hall effect develops fine field structures, which cause 
the field to decay on the Hall time-scale rather than on a much longer ohmic 
time-scale. However, the presence of even a relatively weak poloidal field 
($\sim$ few per cent) changes the result drastically, and dissipation 
proceeds approximately on the ohmic timescale. Since the magnetic field in 
the neutron star should have a rather complex geometry with both the 
poloidal and toroidal components, we can expect that departures in 
the rate of Joule heating caused by the Hall effect have to be small compared 
to the case when the ohmic dissipation is acting alone. Note that some 
simplified models describe the influence of the Hall effect on the magnetic 
evolution in terms of the decay on the Hall timescale (see, e.g., Cumming et 
al. 2004). From a comparison with numerical modelling, it is seen that this 
naive picture is insufficient. The Hall timescale often characterizes the 
period of oscillations rather than the decay timescale of magnetic modes. 
Such an oscillatory behaviour of the decaying magnetic modes was first 
obtained by Urpin \& Shalybkov (1997). Despite Pons \& Geppert (2007) 
having confirmed the earlier finding by Shalybkov \& Urpin (1997) and 
H\"ollerbach \& R\"udiger (2002, 2004) that the Hall effect does not 
substantially accelerate dissipation of realistic magnetic configurations, 
Aguilera, Pons, and Miralles (2007) in their study of Joule heating in 
neutron stars assume confusingly that the Hall currents result in a fast 
decay of the magnetic field. 

The field decay can also be influenced by the so-called Hall-drift 
induced instability if it occurs in neutron stars (Geppert \& Rheinhardt 
2002). These authors found that instability can occur which raises small-scale
magnetic modes if the initial magnetic field is strong and its second 
spatial derivative is large enough. It is unclear, however, whether the effect
of this instability on the dissipation rate is noticeable or not since a 
non-linear regime has not been considered. In particular, Hollerbach \& 
R\"udiger (2002) express doubts that a linear instability starting from small 
amplitude perturbations can be compatible with the cascade caused by the 
non-linear Hall effect. Since even this cascade leads only to small 
departures in the dissipation rate for realistic magnetic configurations 
(Shalybkov \& Urpin 1997, H\"ollerbach \& R\"udiger 2002, 2004), the effect 
of a linear instability should be very weak, if it exists. As far as the 
non-linear regime of instability is concerned, Geppert et al. (2003) assume 
that, at saturation, the amplitude of unstable perturbations can reach a 
significant fraction of the background field and, hence, the latter should 
decay on the Hall timescale. Since this obviously contradicts the observed 
lifetime of pulsars ($\sim 10^8$ yrs), the authors assume that the Hall drift 
induced instability can operate only during part of the neutron star 
life whereas decay before and after such episodes is determined by the ohmic 
dissipation. According to Rheinhardt \& Geppert (2002), episodes occur 
when the age of a neutron star is $\geq 10^{5}$ yrs since the Hall parameter 
is not sufficiently large in younger stars. However, studies of pulsars 
based on methods of population synthesis (Bhattacharya et al. 1992, Hartman 
et al. 1997, Regimbau \& de Freitas Pacheco 2001, Faucher-Giguere \& Kaspi 
2006) indicate that the magnetic field of radiopulsars decays little during 
their active lifetime. A significant magnetic field decay in radiopulsars 
likely can occur only in very young stars with age $\leq 10^{5}$ yrs 
since the number of these pulsars is small and they give a negligible 
contribution to any statistical analysis. Therefore, observations also do not 
support the idea of episodes of the Hall drift induced instability
in neutron stars. Note that recent calculations of the non-linear field 
decay by Pons \& Geppert (2007) confirm the conclusion by Hollerbach \& 
R\"udiger (2002) that, even if the Hall-driven instability exists in neutron 
stars, it has no appreciable influence on the field decay.

\section{Statement of the problem}

To calculate the rate of Joule heating in young neutron stars, we consider 
a very simplified model. We assume that the magnetic field of neutron stars 
is generated by a turbulent dynamo at the beginning of their evolution. The 
dynamo $\alpha$-parameter is proportional to the density gradient and, hence, 
the dynamo is most efficient in the surface layers of proto-neutron stars (see 
Bonanno et al. 2005). Hence, one can expect that there exist electric 
currents in the outer crust just after the convective stage. The decay of 
currents is highly inhomogeneous since the electric resistivity depends on 
the density and varies by many orders of magnitude over the crust. The decay 
timescale is $t_B \approx 4 \pi \sigma L^2/ c^2$, where $\sigma$ is the 
conductivity and $L$ is the lengthscale of the magnetic field, and currents 
deposited closer to the surface will decay faster because the conductivity 
is lower. Therefore, the thermal balance of young neutron stars with age 
$t \leq 10^5$ yrs can be influenced only by dissipation of currents located 
in the surface layers. 

The decay of the crustal magnetic field is governed by the induction 
equation without the convective term
\begin{equation}
\frac{\partial \vec{B}}{\partial t} = - \frac{c^2}{4 \pi} \nabla \times
\left( \hat{R} \nabla \times \vec{B} \right),
\end{equation}
where $\hat{R}= 1/\hat{\sigma}$ is the electric resistivity tensor and 
$\hat{\sigma}$ is the conductivity tensor. In a magnetized crust, components 
of the conductivity tensor are 
\begin{equation}
\sigma_{\parallel} = \sigma_0 \;\;, \sigma_{\perp}
= \frac{\sigma_0}{1 +
\omega_B^2 \tau^2} \;\;, \sigma_{H} = - \frac{\sigma_0 \omega_B \tau}{ 1 +
\omega_B^2 \tau^2} \;\;,
\end{equation}
where $\sigma_{\parallel}$ and $\sigma_{\perp}$ are the components parallel 
and perpendicular to the magnetic field, $\sigma_{H}$ is the so-called Hall 
component that determines the current perpendicular to both the magnetic and 
electric fields, $\sigma_0= e^2 n_e \tau/m_{*}$ is the coductivity at 
$\vec{B} =0$, $n_e$ and $m_{*}$ are the number density of electrons and their
effective mass, $\omega_B$ is the gyrofrequency of electrons, and $\tau$ is
the relaxation time. If the
magnetic field is sufficiently strong and $\omega_B \tau > 1$, then the 
conductivity components across the magnetic field $\sigma_{\perp}$ and
$\sigma_{H}$ are suppressed, and only the conductivity along the magnetic
field is essential. Components of the resistivity tensor $\hat{R} = 1 /
\hat{\sigma}$ are given by
\begin{equation}
R_{\parallel} = R_{\perp} = 1/ \sigma_0 \;\;, R_{H} = \omega_B \tau / 
\sigma_0 = B/ e c n_e \;\;,
\end{equation}
where $n_e$ is the number density of electrons. In a strong magnetic field,
the components parallel and perpendicular to the magnetic field are the same 
and are not suppressed by the magnetic field. The Hall component of 
resistivity $R_{H}$ can be higher than $R_{\parallel}$ if $\omega_B \tau > 1$ 
that, generally, can be fulfilled in some layers of the neutron star.

We neglect the Hall term in our consideration and, strictly speaking, our
results apply to neutron stars with not very strong magnetic fields. However,
as was argued in previous studies (Urpin \& Shalybkov 1997; H\"ollerbach \&
R\"udiger 2004; Pons \& Geppert 2007), even in a very strong magnetic field 
that can magnetize plasma, the Hall effect insignificantly changes the rate 
of dissipation if the magnetic field contains a poloidal component. The 
difference in the rate of dissipation between the model that takes into
account the Hall effect and the hypothetical model that has the same initial
magnetic configuration but evolves in accordance with Eq.~(5) without the 
Hall effect does not exceed 10-20\%. This is much smaller than uncertainties in
our knowlegde of the initial magnetic configuration or the chemical 
composition of the crust. Such accuracy is sufficient for our purposes, and
we will mimic the field decay by Eq.~(5).  
    
We restrict our consideration to a dipolar field that can be described by
the vector potential $\vec{A}= (0, 0, S(r, t) \sin \theta /r)$, where $r$ and 
$\theta$ are the radius and polar angle, respectively. The function 
$S(r, t)$ obeys the equation
\begin{equation}
\frac{\partial^2 S}{\partial r^2} - \frac{2 S}{r^2} = \frac{4 \pi 
\sigma_0}{c^2}
\frac{\partial S}{\partial t},
\end{equation}
with the boundary conditions $\partial S/\partial r + S/R = 0$
at the stellar surface $r=R$ and $S \rightarrow 0$ in the deep crustal 
layers. The $\varphi$-component of the electric current maintaining the 
dipolar magnetic field is 
\begin{equation}
j_{\varphi}= - \frac{c}{4 \pi} \frac{\sin \theta}{r} \left( 
\frac{\partial^2 S}{\partial r^2} - \frac{2 S}{r^2} \right).
\end{equation}
We have for the rate of Joule heating $\dot{q}(r, \theta, t) = j_{\varphi}^2/
\sigma_0$. For simplicity, nonsphericity is often neglected in cooling 
calculations. Therefore, we will use the polar-averaged expression for 
the rate of Joule heating. It is also convenient to normalize the function 
$S(r, t)$ to its initial value at the surface, $S(R, 0)$, which in turn can 
be related to the initial field strength at the magnetic equator $B_e$ by 
$S(R, O) = R^2 B_e$. Then, the polar-averaged rate of Joule heating is given 
by 
\begin{equation}
\dot{q} = \frac{c^2 R^4 B_e^2}{24 \pi^2 r^2 \sigma_0}
\left( \frac{\partial^2 s}{\partial r^2} - \frac{2 s}{r^2} \right)^2.
\end{equation}
In young neutron stars with age $t < 10^5$ yrs, the conductivity of the 
outer crust and the outer part of the inner crust is determined by electron 
scattering on phonons. Scattering on impurities is important only in deep 
layers of the inner crust even if the impurity parameter is large (see, e.g., 
Jones 2004). We use the conductivity due to electron-phonon scattering 
calculated by Itoh et al. (1993). These conductivities depend on the
temperature. Calculations presented here are based on the standard
cooling scenario, which corresponds to a star with standard neutrino 
emissivities. We consider the field decay using the thermal history for the
1.4 $M_{\odot}$ neutron star constructed with the equation of state
of Friedman \& Pandharipande (1981) which is representative of intermediate
equations of state. Our choice is compelled by the fact that models
with the equation of state stiffer than that of Friedman \& Pandharipande
agree better with data on the magnetic evolution of pulsars
(Urpin \& Konenkov 1997). 

Calculating the rate of Joule heating, we neglect its back influence on the 
thermal evolution of a neutron star. It is not the purpose of this paper to
discuss in detail how the thermal and magnetic evolution are coupled. One 
can neglect this coupling if the rate of heating does not exceed the 
luminosity of a standard neutron star. Our approach allows us to obtain the 
lower limit on the field strength that can 
affect the thermal evolution due to Joule heating, and we compare this limit 
to observational data. Note, however, that even the temperature of magnetars
which are believed to be heated by the decay of extremely strong fields $\geq 
10^{14}$ G is only a factor $\sim 2-3$ higher than the temperature of a 
standard cooling neutron star. Therefore, our calculations can provide an 
order of magnitude estimate for the rate of Joule heating in the case of 
such strong magnetic fields. This simplified approach can be useful in 
understanding the main qualitative features of Joule heating in magnetars. 
A study of the coupled early magnetothermal evolution of very strongly 
magnetized stars such as magnetars is a more complicated problem, and it will 
be considered in a forthcoming paper. 

If the field in the surface layers is very strong, such as $B^2/4 \pi > \mu$ 
where $\mu$ is the shear modulus of the crust, then the induction and 
momentum equations are coupled, and magnetic stresses can induce motion 
of the crustal material. Therefore, decay of 
the crustal magnetic field can change the Lorentz force and, generally, can 
lead to crustal motion. The shear modulus is $\mu \sim 0.1 n_i (Ze)^2 / a$ 
where $a= (3/4 \pi n_ị)^{1/3}$, $Z$ and $n_i$ is the charge and 
number density of ions (Strohmayer et al. 1991). The condition $B^2 /4 \pi > 
\mu$ is approximately equivalent to $\rho < 10^{9} B_{13}^{3/2}$ g/cm$^3$ 
where $B_{13} =B/ 10^{13}$ G, and one should expect that Joule heating of the
surface layers in high magnetic field pulsars is accompanied by motion in 
the crust. We will neglect, however, the contribution of this effect to
heating.

\section{Numerical results}

In calculations, we assume that the magnetic field generated by a dynamo 
initially occupies the outer layer of the crust up to a density $\rho_0$ thus, 
initially, $B \neq 0$ at $\rho \leq \rho_0$. Since the field is confined
to the crust, the density $\rho_0$ can be lower or equal to the density at 
the crust-core boundary, $\sim 2 \times 10^{14}$ g/cm$^3$. It is  
difficult to determine the inner boundary of the generation region from the 
physical data since both turbulent dynamo coefficients and stability 
properties of proto-neutron stars are subject to many uncertainties. 
Therefore, we perform calculations for a few different values of $\rho_0$. 
Note that the magnetic evolution is in good agreement with pulsar data if 
$\rho_0 \sim 10^{13}$ g/cm$^3$ (Urpin \& Konenkov 1997). This conclusion is 
in qualitative agreement with the dynamo models of neutron stars since the 
dynamo is efficient only in the surface layers where the density gradient is 
large. After the dynamo action stops in a proto-neutron star, the field 
diffuses into deeper layers and can be non-vanishing at $\rho > \rho_0$. 
The initial distribution of the magnetic field at $\rho < \rho_0$ is also 
uncertain but, fortunately, the main qualitative results are not crucially 
sensitive to this distribution. We consider two models of the initial 
distribution shown in Fig.~1. These models differ by the characteristic 
depth of currents with model 2 (shown in the dashed line) corresponding 
to a smaller depth than model 1 (solid line).
 
\begin{figure}
\begin{center}
\includegraphics[width=8.0cm, angle=270]{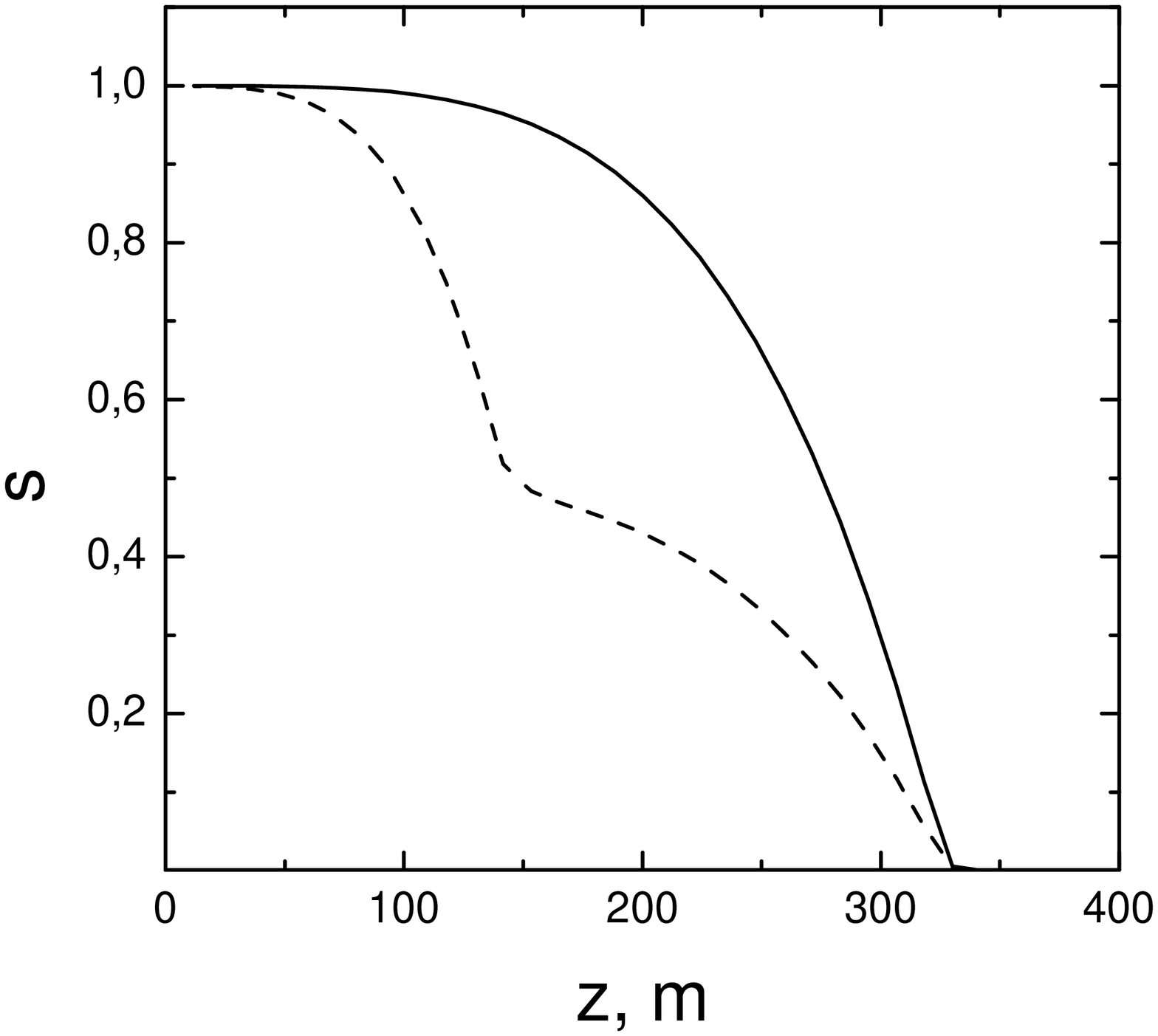}
\caption{The dependence of the initial distribution of $s(0, r)$ on the depth
from the surface $z=R-r$ for the models 1 (solid line) and 2 (dashed line) and for $\rho_0 = 10^{13}$ g/cm$^3$ (the corresponding depth $z_0$ is 
approximately 330 m.}
\end{center}
\end{figure}

In Fig.~2, we plot the distribution of the rate of Joule heating, $\dot{q}$, 
for different ages. The distribution turns out to be complicated 
with a deep minimum that originates because, for a dipole field, the current 
density changes sign at some depth. The depth of this minimum increases 
with time due to diffusion of the magnetic field into deep layers. The origin
of this minimum is caused by the general diffusive properties of the crustal
magnetic field and can be understood from the simplest qualitative 
consideration. Indeed, if the initial distribution of $s$ is like those shown 
in Fig.1, then the magnetic field will diffuse with time into deep layers 
with $\rho > \rho_0$ and, hence, the time derivative of $s$ will be positive 
somewhere at $\rho \geq \rho_0$, at least. On the another hand, the value of 
$s$ and the field strength at the surface will obviously decrease with time 
because of ohmic dissipation and,hence, we have $\partial s/\partial t < 0$ 
at $r=R$. Since $s$ is a continuous function of the radius $r$ and $\partial 
s/\partial t$ has different signs at the surface and somewhere in deep layers,
there always exists at least one point inside the neutron star crust where 
$\partial s/ \partial t=0$. Combining Eqs.~(8) and (9), we find that 
$j_{\varphi} = 0$ at that point as well. Since $\dot{q} \propto 
j_{\varphi}^2$, the rate of heating is also vanishing at the same depth. In 
Fig.~2, we do not show $\dot{q}$ at this depth in detail because the region 
where $\dot{q} \approx 0$ is very narrow. The first maximum corresponds to a 
relatively low density $\sim 10^{10} - 10^{12}$ g/cm$^3$, depending on the 
model. Initially, this maximum is located at a very low density, but it moves 
slowly inward with the time. After $10^4$ yrs, the maximum reaches a depth 
$\sim 300$ m corresponding to $\rho \sim 10^{12}$ g/cm$^3$. At that time, the 
distribution of $\dot{q}$ is already rather flat in the surface layers. The 
second maximum of $\dot{q}$ is located substantially deeper, at densities 
$\sim 3 \times 10^{12} - 3 \times 10^{13}$ g/cm$^3$, and it also moves inward 
with time. The position of the second maximum is less sensitive to the 
initial models because our models 1 and 2 differ mainly in the distribution 
of currents in the surface layers. Perhaps such a complicated distribution 
of $\dot{q}$ can result in a complicated radial dependence of the temperature 
in strongly magnetized stars. The rate of heating can reach a very high value 
$\sim 10^{16}-10^{18}$ erg/cm$^3 \cdot$s or even higher if the initial 
magnetic field is stronger than $B \geq 3 \times 10^{13}$ G. The total rates 
of heating above and below the region where $\dot{q} \approx 0$ are of the 
same order of magnitude.

\begin{figure}
\begin{center}
\includegraphics[width=9.0cm, angle=270]{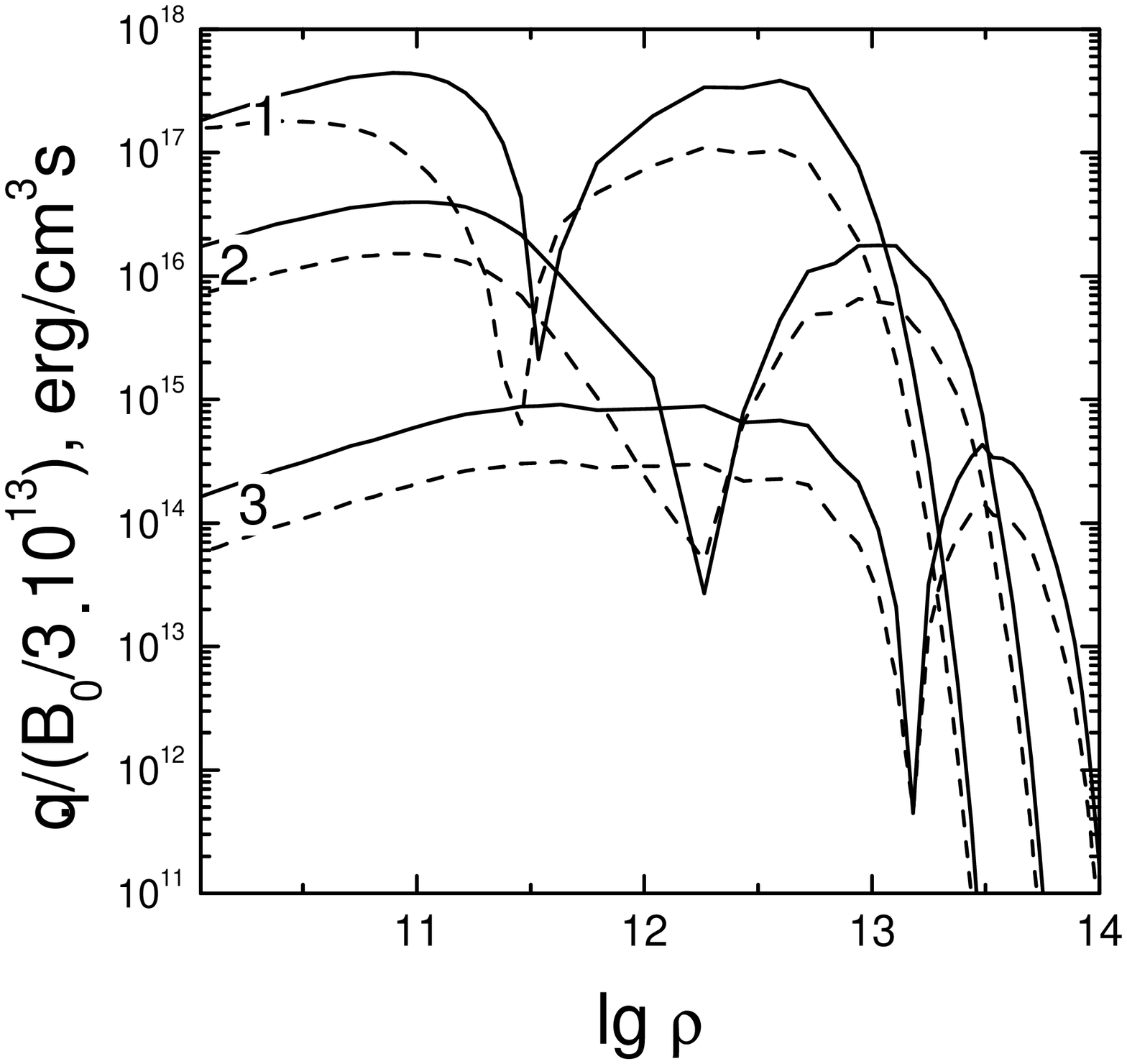}
\caption{The dependence $\dot{q}$ on $r$ for different ages and models.
The solid and dashed lines corresponds to the models shown in the solid 
and dashed lines in Fig.~1. Curves 1, 2, and 3 show the distribution for 
$t=10^2$, $10^3$, and $10^4$ yrs, respectively.}
\end{center}
\end{figure}

In Fig.~3, we plot the evolution of the density $\rho_{*}$ where the rate 
of Joule heating is maximal for models 1 and 2. This density corresponds 
to the first maximum of $\dot{q}$ (see Fig.~2). The location of the layer 
where the maxumum energy is released is important for understanding the energy
budget in neutron stars (see, e.g., Kaminker et al. 2006). During the initial 
$\sim 10^3$ yrs, the depth of this layer is very small and corresponds to 
$\rho_{*} < 10^{11}$ g/cm$^3$. At this stage, $\rho_{*}$ depends on the 
initial distribution of currents. Later on, at $t > 10^3$ yrs, this depth 
becomes almost independent of the initial distribution. Up to $t \sim 10^4$ 
yrs, the maximum of Joule heating is located in the layers with a density 
lower than $4 \times 10^{11}$ g/cm$^3$. Likely, heat released in the layers 
with such a low density will be transfered to the surface and emitted by 
photons rather than diffusing to the core and being emitted by neutrinos.

\begin{figure}
\begin{center}
\includegraphics[width=9.0cm, angle=270]{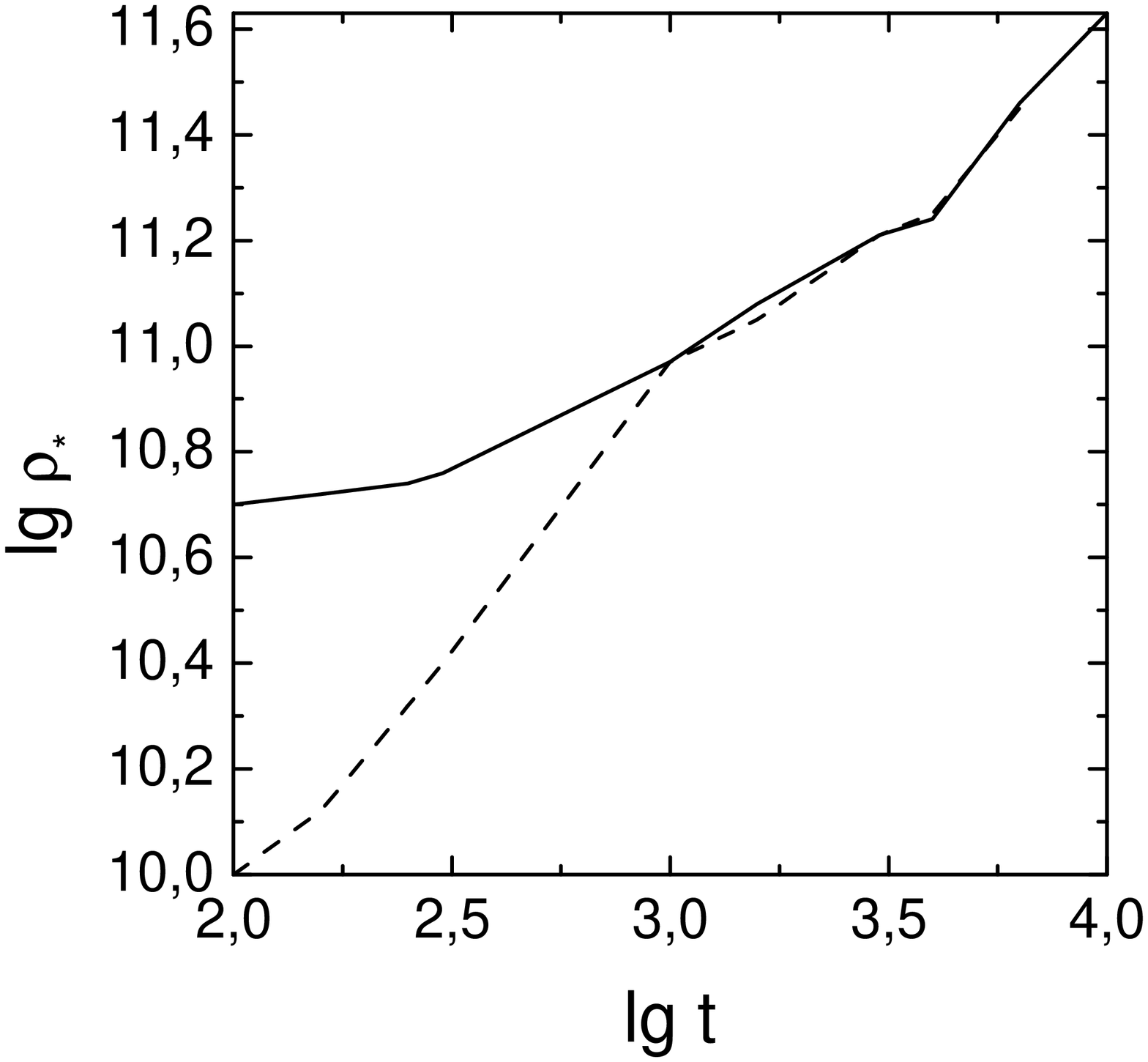}
\caption{The depth of the first maximum of $\dot{q}$ as a function of 
time for the models shown in Fig.~1.}
\end{center}
\end{figure}

In Fig.~4, we show the evolution of the total rate of Joule heating integrated
over the crust volume, $\dot{Q} = \int \dot{q} dV$. For comparison, we also
plot in this figure the photon luminosity of a neutron star with standard
cooling. If the initial magnetic field is strong and the corresponding heating
rate is above the dotted line, our results can provide only an order of 
magnitude estimate because Joule heating will influence the neutron star 
thermal evolution. The rate of Joule heating is calculated for three values of
$\rho_0 = 10^{12}$, $10^{13}$, and $10^{14}$ g/cm$^3$. For all these initial 
distributions of the magnetic field, the rate of Joule heating can reach 
rather high values $\sim 10^{34}-10^{35}$ erg/s during $\sim 10^3$ yrs if the 
initial field strength is $3 \times 10^{13}$ G. This heating rate is higher 
than the luminosity of a standard neutron star without any additional heating 
mechanisms. At a given initial field strength, the heating rate is initially 
higher for smaller $\rho_0$, but it decreases faster with the time. 
For example, if the magnetic field initially occupies the surface layer with 
a density $\rho < \rho_0 = 10^{12}$ g/cm$^3$, then the heating rate at $t \sim
100$ yrs is approximately a factor of 10 higher than in the case of $\rho_0 
=10^{14}$ g/cm$^3$. However, the heating rate for $\rho_0 = 10^{12}$ g/cm$^3$ 
decreases very rapidly and becomes smaller already after $\sim 2 \times 10^3$ 
yrs. Generally, the rate of Joule heating can be high in young neutron
stars. Even in pulsars with initial magnetic fields $\sim 3 \times 10^{13}$, 
Joule heating exceeds the photon luminosity during $\sim 10^4$ yrs if $\rho_0 
\sim 10^{13}$ g/cm$^3$. Therefore, young pulsars with such magnetic fields 
should be hotter than it is predicted by the standard cooling scenario. High 
magnetic field pulsars such as PSR J1847-0130 ($B \approx 9 \times 10^{13}$ G)
or PSR J1718-3718 ($B \approx 7 \times 10^{13})$ can be in such a ``heated'' 
state even longer ($\sim 10^5$ yrs). However, after that time, their thermal 
evolution will follow the standard scenario. Our calculations show that even 
the rate of heating required for magnetars, $\dot{Q} \sim 10^{35} - 3 \times 
10^{36}$ erg/s, is reachable during the initial evolution if the initial
magnetic field at the surface is $\sim 10^{14}$ G. Note that the rate of 
Joule heating depends not only on the field strength but also on the depth of 
the region where the magnetic field was generated. This dependence is rather 
sensitive and, generally, it is possible that the heating rate differs by 
more than an order of magnitude in pulsars of the same age and having 
approximately the same surface magnetic fields (for example, compare the 
solid and dash-and-dotted lines in Fig.~4). Such different heating rates can 
lead to a substantial difference in the surface temperature in high magnetic 
field pulsars.

\begin{figure}
\begin{center}
\includegraphics[width=9.0cm, angle=270]{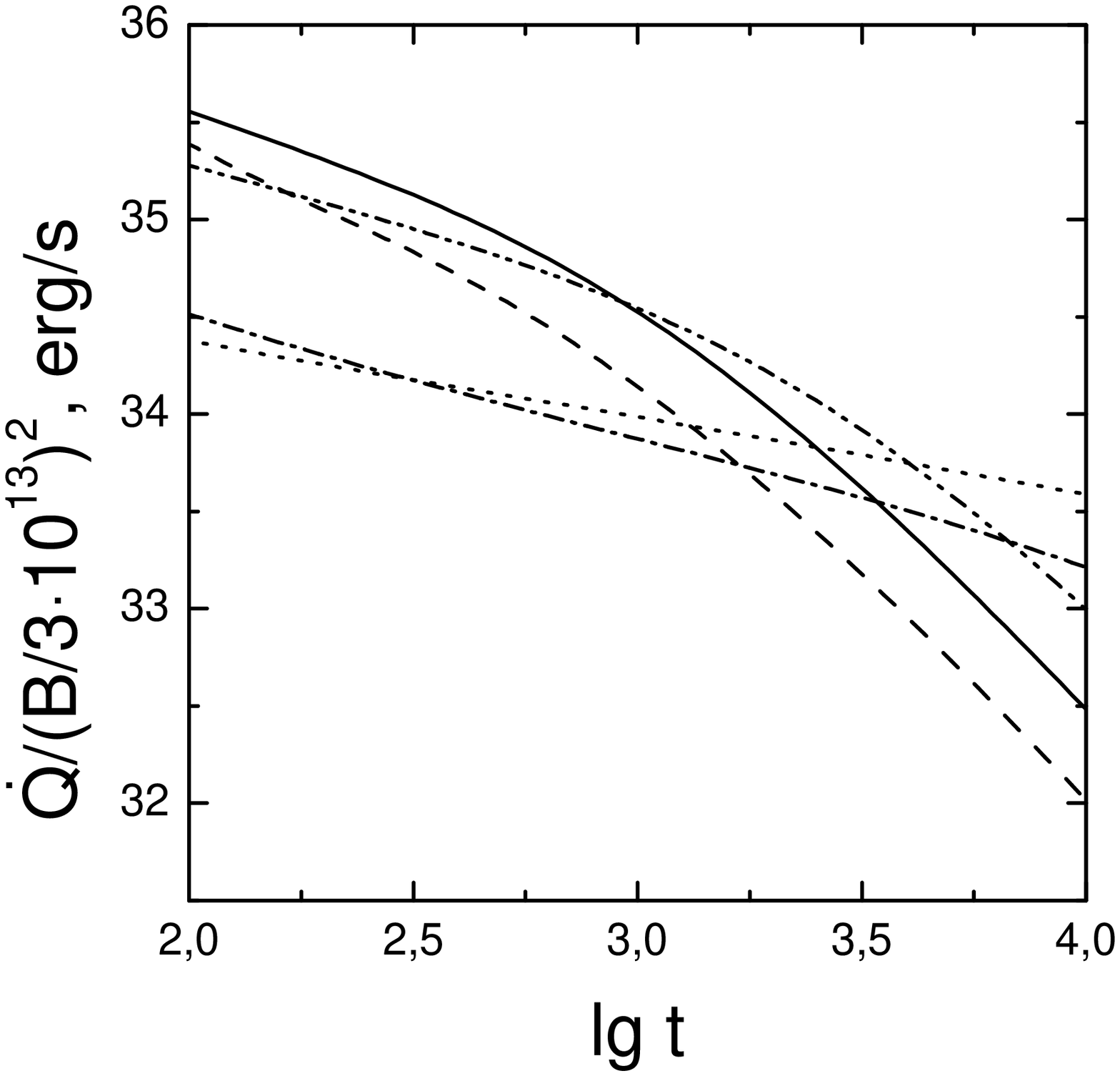}
\caption{The dependence of $\dot{Q}$ on time for different models. The
magnetic field is normalized to $3 \times 10^{13}$ G. The solid and dashed
lines correspond to the initial models 1 and 2 from Fid.~1, respectively, 
with $\rho_0 = 10^{12}$ g/cm$^3$. The dashed-and-dotted and dotted-and-dashed
lines show $\dot{Q}$ for the model 1 with $\rho_0 = 10^{14}$ and $10^{13}$
g/cm$^3$, respectively. The dotted line shows the time 
dependence of the luminisity of a neutron star with no additional heating 
mechanism.}
\end{center}
\end{figure}

\section{Discussion}

We have considered dissipation of a magnetic field initially anchored to 
a neutron star crust. The magnetic field of neutron stars can be generated 
by a turbulent dynamo during a short hydrodynamically unstable phase 
at the beginning of their evolution. The mean-field dynamo is most effecient 
in the surface layers where the density gradient is large and, hence, the 
dynamo $\alpha$-parameter is maximal. Therefore, electric currents maintaining
the magnetic configuration are likely located in the surface layers. 
Dissipation of electric currents is a highly inhomogeneous process because the
density and, hence, conductivity varies by many orders of magnitude through 
the crust. The rate of dissipation decreases with increasing density and,
hence, the field decay is slower in deep crustal layers. Dissipation of 
currents located in the surface layers proceeds on a relatively short 
time-scale and is accompanied by massive heat release. Joule heating is so 
efficient that it can even influence the thermal evolution of young neutron 
stars if the magnetic field is sufficiently strong. 

The duration of the stage when Joule heating can affect the thermal evolution
is determined by the strength of the magnetic field and the depth where this 
field is anchored. Both these quantities can vary over a wide range because 
they depend, for example, on poorly known details of the generation process 
and post-collapse hypercritical accretion that can essentially submerge the 
generated magnetic field (Chevalier 1989). At a given field strength, the rate
of Joule heating is higher in a star where the field was initially anchored in
the layer with a lower density. However, the time interval when this heating 
is important for the thermal evolution is shorter in such stars. On the 
contrary, if the field is anchored in a deeper layer, Joule heating is less 
effecient but it can contribute significantly in the thermal evolution over 
longer time. 

Since the rate of Joule heating depends on two parameters it is generally 
possible that neutron stars with the same age and surface magnetic 
field can have substantially different heating rates and, hence, surface 
temperatures. This allows us to understand why the surface temperature of the 
high magnetic field pulsar PSR J1718-3718 is substantially lower than that 
of magnetars despite its magnetic field being approximately that of magnetars
($7.4 \times 10^{13}$). The magnetic field of high magnetic field pulsars 
was likely anchored initially in layers with a lower density, and the 
stage of efficient Joule heating when their luminosity was comparable to 
that of magnetars is completed.
 
Our calculations show that additional heating provided by dissipation of the 
crustal magnetic field can even be comparable to the luminosity of magnetars 
($\sim 10^{35}-10^{37}$ erg/s) if the magnetic field at the stellar surface 
is as high as $\sim (1-3) \times 10^{14}$ G. Generally, such a strong magnetic
field can be generated in rapidly rotating proto-neutron stars by the 
turbulent $\alpha$-dynamo (Bonanno et al. 2006). Note that our calculations
provide only the lower limit on the rate of Joule heating since we neglected
the back influence of this heating on the crustal temperature. Additional 
heating will increase the temperature and decrease the conductivity and, 
hence, it will result in a higher dissipation rate if one consistently takes 
into account the coupling between the thermal evolution and field decay. 
For strongly magnetized stars with a surface field $\sim (1-3)
\times 10^{14}$ G, the epoch when Joule heating is of the order of the 
magnetar luminosity can last as long as $\sim 10^4-10^{5}$ yrs if the magnetic
field was initially anchored in the layers with $\rho_0 \sim 10^{13}$ 
g/cm$^3$. This time is comparable to the estimated life-time of magnetars. If 
the initial field is weaker. $\sim 3 \times 10^{13}$ G, but is confined to the
same layers, then the rate of Joule heating exceeds the standard luminosity of
a cooling neutron star for a shorter time, $\sim 10^3$ yrs. During this 
time, the rate of heating is approximatelly one order of magnitude greater 
than the luminosity of a standard neutron star with the same age. Even in 
neutron stars born with a relatively weak magnetic field $\sim 3 \times 
10^{12}$ G, Joule heating can influence the early thermal elolution but 
for a very short time and only if the magnetic field was initially confined
to low density layers.  

From our results, we conclude that many young neutron stars can 
ehxibit departures from the standard cooling scenario if the magnetic field
is of crustal origin and if it is sufficiently strong. How large these 
departures are and how long Joule heating can manifest  is determined 
by the initial strength of the magnetic field and its location. At the moment,
observations only provide information on the surface temperature of neutron 
stars with an age greater than $\sim 1$ kyr. For stars with such age, our 
model predicts that departures from standard cooling can be expected only 
if the magnetic field is sufficiently strong, $B > (1-3) \times 10^{13}$ G. 
Therefore, recently discovered high magnetic field pulsars
such as PSR J1847-0130 ($9.4 \times 10^{13}$ G), PSR J1718-3718 ($7.4 \times 
10^{13}$ G), PSR J1814-1744 ($5.5 \times 10^{13}$ G), PSR J1119-6127 
($4.4 \times 10^{13}$ G), and PSR B0154+61 ($2.1 \times 10^{13}$ G) all
can have a higher surface temperature than is predicted by standard
cooling models without additional heating.

\begin{figure}
\begin{center}
\includegraphics[width=9.0cm, angle=270]{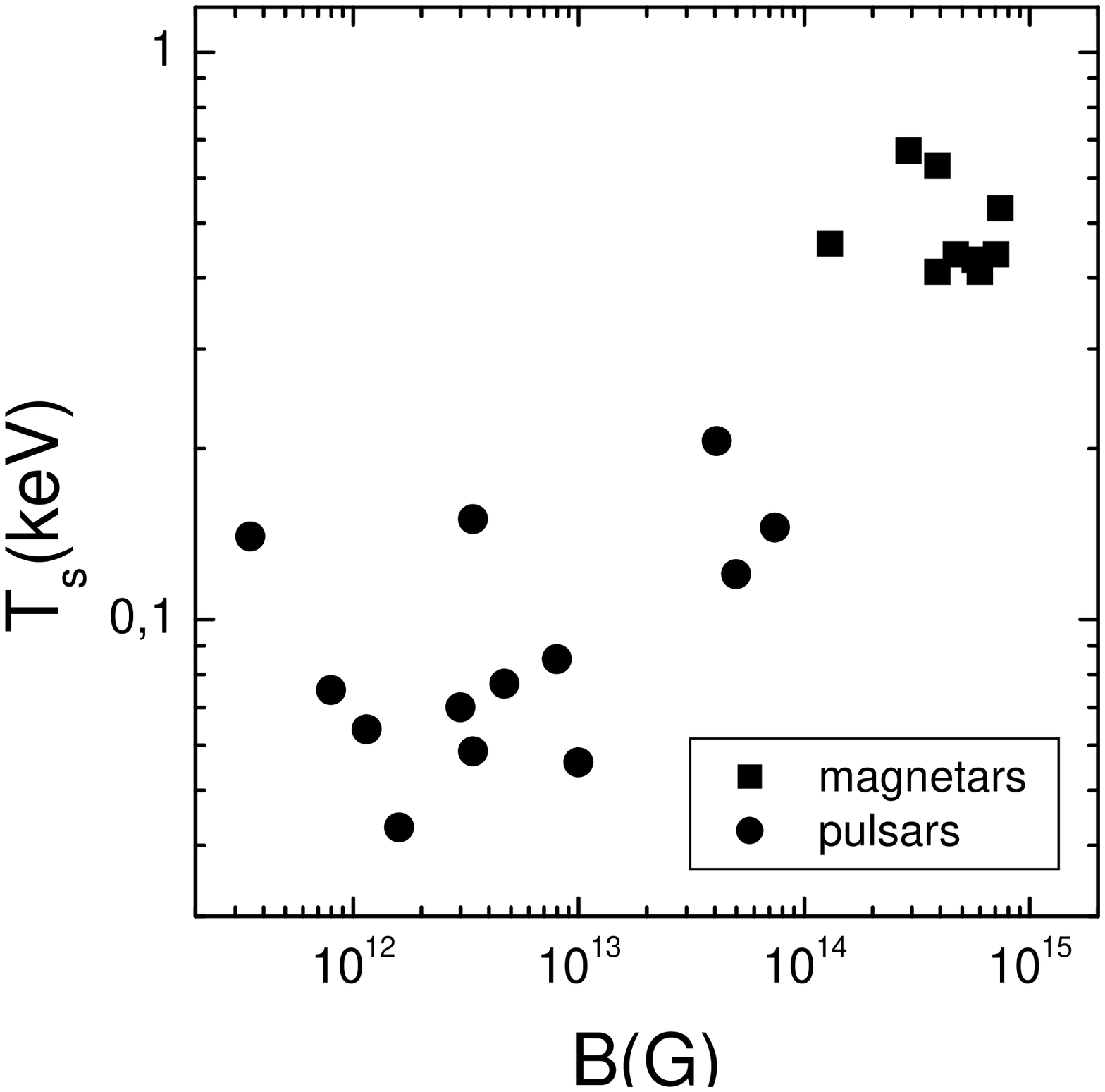}
\caption{The surface temperature $T_s$ versus the magnetic field for
22 young neutron stars.}
\end{center}
\end{figure}

In Fig.~5, we plot the available data on the surface temperature $T_s$ and 
magnetic fields $B$ of 22 young neutron stars. These data include 9 magnetars
listed in Table 1 by Pons et al. (2007) which are likely young objects
according to current theory. We also plot the data on $B$
and $T_s$ for 13 radiopulsars with a spin-down age shorter than 1 Myr. If 
the magnetic fields of these pulsars decay, their true age should be 
substantially less than the spin-down age and, likely, such pulsars are 
sufficiently young to be treated in our analysis. These 13 radiopulsars are 
listed in Table 1.  
  
\begin{table}
\begin{center}
Table 1. \\
The surface temperature and magnetic field of young radiopulsars. \\
\end{center}
\begin{center}
\begin{tabular}{|l|l|l|l|}
\hline
Source   &  $kT$   & $B$               &  Ref.      \\
         &  (keV)  & (10$^{12}$ G)     &            \\
\hline
       &          &               &                 \\
RX J0822-4320  &  0.14-0.16  & 3.4  & Zavlin et al.(1999)  \\
               &             &      & Pavlov et al.(1999)  \\     
1E 1207.4-5209 &  0.12-0.16  & 0.35 & Zavlin et al.(2004)  \\
               &             &      & Gotthelf et al.(2007) \\
Vela           &  0.06-0.07  & 3.4  & Page et al.(2004)    \\
PSR B1706-44   &    0.07     & 3.0  & Page et al.(2004)    \\
PSR J0538+2817 &    0.075    & 0.8  & Levandowski et al.   \\
               &             &      & (2004)               \\
Geminga        &  0.043      & 1.6  & De Luca et al.(2005) \\
PSR 1055-52    &   0.064     & 1.15 & De Luca et al.(2005) \\
PSR B0656+14   &   0.077     & 4.7  & De Luca et al.(2005) \\
PSR B2334+61   &   0.056     & 10   & McGowan et al. \\
               &             &      & (2006)               \\
PSR J1718-3718 &   0.145     & 74   & Kaspi at al.(2005)   \\
PSR J1119-6127 &   0.206     & 41   & Gonzalez et al.(2005) \\
CXOU J1819-1458 &  0.120     & 50   & Reynolds et al.(2006) \\
PSR J1357-6429 &   0.085     & 8    & Zavlin (2007)         \\
       &          &               &                \\
\hline
\end{tabular}
\end{center}
\end{table}

In accordance with our calculations, the thermal history of young pulsars 
with an age $\leq 10^5$ yrs can be influenced by Joule heating if the 
magnetic field at birth is stronger than $(1-3) \times 10^{13}$ G. Indeed,
it seems that the surface temperature of high magnetic field pulsars with $B >
10^{13}$ G shows some trend to be higher than that of low 
magnetic field pulsars, in qualitative agreement with the prediction of 
our model. This difference cannot be accounted for by the influence of the 
magnetic field on the thermal conductivity, because the geometry and strength
of the magnetic field are unimportant for the surface averaged thermal 
structure of neutron stars if the field strength is moderate, $B \leq 10^{14}$
G (Potekhin et al. 2005). For such $B$, the magnetic field can affect the
thermal evolution of neutron stars only via Joule heating, and the surface 
temperature should be proportional to approximately $B^{1/2}$. The data in 
Fig.~5 obviously are not sufficient to prove this dependence but, at least, 
they do not contradict it. More detections of the surface temperature in high 
magnetic field pulsars are needed, and they could provide strong evidence 
of the crustal origin of the magnetic field in neutron star.

It is important to consider coupled magnetothermal evolution by solving the 
heat and induction equations because characteristic timescales of the thermal 
and crustal magnetic evolution are approximately comparable in young neutron 
stars with $t < 10^5$ yrs. However, this problem, taking into account the 
magnetically-induced anisotropy of the thermal conductivity and electric 
resistivity tensors,is far from being solved and the study of particular 
problems of the magnetothermal evolution can help in understanding when Joule 
heating can affect neutron star cooling. From the result of the present study 
and conclusions obtained by Miralles et al. (1998), Joule heating can be 
important only in very young ($t < 10^4 - 10^5$ yrs) or in old ($t > 10$ Myr) 
neutron stars provided the magnetic field is sufficiently strong at birth. 
In stars with a moderate age, Joule heating seems to be inefficient.

\section*{Acknowledgments}
This work has been supported by the Spanish Ministerio de Ciencia y 
Tecnologia grant AYA 2001-3490-C02. VU thanks Generalitat Valenciana and
Universitat d'Alacant for a financial support and hospitality.


\begin{thebibliography}{}

\bibitem{}
Aguilera, D., Pons, J., \& Miralles J.A. 2008, ApJ, 673, L167



\bibitem{}
Bhattacharya, D., Wijers, R., Hartman, J., \& Verbunt, F. 1992, A\&A, 254, 198

\bibitem{}
Braithwaite, J. \& Nordlund, A. 2006, A\&A, 450, 1077

\bibitem{}
Braithwaite, J. \& Spruit, H. 2006, A\&A, 450, 1097

\bibitem{}
Bonanno A., Rezzolla L., Urpin V. 2003. A\&A, 410, L33

\bibitem{}
Bonanno, A., Urpin, V., \& Belvedere, G. 2005, A\&A, 440, 199

\bibitem{}
Bonanno, A., Urpin, V., \& Belvedere, G. 2006, A\&A, 451, 1049

\bibitem{}
Chevalier, R. 1989, ApJ, 346, 847

\bibitem{}
Cumming, A., Arras, P., \& Zweibel, E. 2004, ApJ, 609, 999

\bibitem{}
De Luca, A., Caraveo, P., Mereghetti, S., Negroni, M., \& Bignami, G.
2005, ApJ, 623, 1051 

\bibitem{}
Faucher-Giguere, C.-A. \& Kaspi, V. 2006, ApJ, 643, 332

\bibitem{}
Friedman B., Pandharipande V. 1981, Nucl. Phys. A, 361, 502

\bibitem{}
Geppert, U. \& Rheinhardt, M. 2002, A\&A, 392, 1015

\bibitem{}
Geppert, U., Rheinhardt, M., \& Gil, J. 2003, A\&A, 412, L33

\bibitem{}
Gonzalez, M., Kaspi, V., Camilo, F., Gaensler, B., \& Pivovaroff, M. 2005,
ApJ, 630, 489

\bibitem{}
Gotthelf,E. \& Halpern, J. 2007, ApJ, 664, L35

\bibitem{}
Haensel, P. \& Zdunik, J. 1990, A\&A, 227, 431

\bibitem{}
Hartman, J., Bhattacharya, D., Wijers, R., Verbunt, F. 1997, A\&A, 322, 477

\bibitem{}
Hollerbach, R. \& R\"udiger, G. 2002, MNRAS, 337, 216

\bibitem{}
Hollerbach, R. \& R\"udiger, G. 2004, MNRAS, 347, 1273

\bibitem{}
Itoh, N., Hayashi, H., \& Kohyama, Y. 1993, ApJ, 418, 405

\bibitem{}
Jones, P.B. 2004, Phys. Rev. Lett., 93, 1101 

\bibitem{}
Kaminker, A., Yakovlev, D., Potekhin, A., Shibazaki, N., Shternin, P.,
\& Gnedin, O. 2006, MNRAS, 371, 477

\bibitem{}
Kaspi, V. \& McLaughlin, M. 2005, ApJ, 618, L41

\bibitem{}
Konar, S. 2002, MNRAS, 333, 475

\bibitem{}
Lewandowski, W., Wolszczan, A., Feiler, G., Konacki, M., \& Soltysinski, T.
2004, ApJ, 600, 905

\bibitem{}
McGowan, K., Zane, S., Cropper, M., Vestrand, W., \& Ho, C. 2006, ApJ, 639, 
377

\bibitem{}
Miralles, J.A., Urpin, V., \& Konenkov D. 1998, ApJ, 503, 368

\bibitem{}
Mitra, D., Konar, S., \& Bhattacharya, D. 1999, MNRAS, 307, 459

\bibitem{}
Muslimov, A. 1994, MNRAS, 267, 523

\bibitem{}
Naito, T. \& Kojima, Y. 1994, MNRAS, 266, 597

\bibitem{}
Negele J. \& Vautherin D. 1973, Nucl. Phys. A207, 298

\bibitem{}
Page, D., Geppert, U., \& Zannias T. 2000, A\&A, 360, 1052

\bibitem{}
Page, D., Lattimer, J., Prakash, M., \& Steiner, A. 2004, ApJS, 155, 623

\bibitem{}
Pavlov, G., Zavlin, V., \& Truemper, J. 1999, ApJ, 511, 45

\bibitem{}
Perez-Azorin, J.F., Miralles, J.A., \& Pons J. 2006, A\&A, 451, 1009 


\bibitem{}
Pons, J. \& Geppert, U. 2007, A\&A, 470, 303

\bibitem{}
Pons, J., Link, B., Miralles, J., \& Geppert, U. 2007, Phys. Rev. L, 98, 1101

\bibitem{}
Potekhin, A., Yakovlev, D., Chabrier, G., \& Gnedin O. 2003, ApJ, 594, 404

\bibitem{}
Potekhin, A., Urpin, V., \& Chabrier, G. 2005, A\&A, 443, 1025

\bibitem{}
Regimbau, T. \& de Freitas Pacheco, J. A. 2001, A\&A, 374, 182

\bibitem{}
Reynolds, S., et al. 2006, ApJ, 639, L71

\bibitem{}
Rheinhardt, M. \& Geppert, U. 2002, Phys. Rev. Lett., 88, 1103

\bibitem{}
R\"udiger, G. \& Kitchatinov L. 1993, A\&A, 269, 581

\bibitem{}
Sengupta, S. 1998, ApJ, 501, 792

\bibitem{}
Shalybkov, D. \& Urpin, V. 1997, 321, 685

\bibitem{}
Slattery, V., Doolen, G., \& De Witt, H. 1980, Phys. Rev. A, 21, 2087

\bibitem{}
Strohmayer, T., Van Horn, H.M., Ogata, S., Iyetomi, H., \& Ichimaru, S.
1991, ApJ, 375, 679

\bibitem{}
Thompson, C. \& Duncan, R. 1993, ApJ, 408, 194


\bibitem{}
Urpin, V. \& Muslimov, A. 1992. MNRAS, 256, 261

\bibitem{}
Urpin, V. \& Gil, J. 2004, A\&A, 415, 305 

\bibitem{}
Urpin, V. \& Konenkov, D. 1997, MNRAS, 292, 167

\bibitem{}
Zavlin, V. 2007, ApJ. 665, L143 

\bibitem{}
Zavlin, V., Pavlov, G., \& Sanwal, D. 2004, ApJ, 606, 444

\bibitem{}
Zavlin, V., Truemper, J., \& Pavlov, G. 1999, ApJ, 525, 959

\end{thebibliography}
\end{document}